\documentstyle[prl,aps,twocolumn]{revtex}
\tighten\draft
\begin{document}   

\title{A regular $C^0$ singularity is not necessarily weak.}
\author{Brien C. Nolan\footnote{e-mail: nolanb@ccmail.dcu.ie}}
\address{
School of Mathematical Sciences, Dublin City University, Glasnevin, Dublin 9, Ireland}
\date{\today}

\maketitle

\begin{abstract}

Examples of space-times are given which contain scalar curvature
singularities whereat the metric tensor is regular and continuous,
but which are gravitationally strong. Thus the argument that such
singularities are necessarily weak is incomplete; in particular
the question of the gravitational strength of the null Cauchy
horizon singularity which occurs in gravitational collapse remains
open.
\newline
\pacs{PACS: 04.20.Dw, 04.70.-s, 04.20.Jb}
\end{abstract}



\newcommand{\be}{\begin{eqnarray}}
\newcommand{\ee}{\end{eqnarray}}


The picture of the internal structure of black holes formed by
gravitational collapse has been greatly revised over the last
decade \cite{haifa}. The BKL \cite{BKL} scenario which involves a
crushing, space-like, oscillatory singularity has been replaced by
a more subtle picture involving the null mass-inflation Cauchy
horizon (CH) singularity which precedes a strong space-like
central singularity \cite{PI,ori,BDM,HP}. It has been argued
repeatedly that the null CH singularity is gravitationally weak;
that is, physical objects only suffer a finite tidal distortion on
encountering this singularity. An appropriate mathematical model
of this was given by Ellis and Schmidt \cite{ES} and by Tipler
\cite{tipler}. It is worth repeating the definitions here.

Consider a causal geodesic $\gamma:[t_0,0)\rightarrow M$ in a
space-time $(M,g)$ which approaches a singularity as $t\rightarrow
0$. For any $t_1\in[t_0,0)$, let $J_{t_1}(t)$ be the set of Jacobi
fields orthogonal to and along $\gamma$ (at parameter value $t$)
which vanish at $t=t_1$. Then the exterior product of any three
(for time-like geodesics) or two (for null geodesics) independent
Jacobi fields forms a volume element $V(t)$ along $\gamma$. The
geodesic is said to terminate in a strong (weak) singularity if
$\|V(t)\|$ is zero or infinite (non-zero and finite) in the limit
$t\rightarrow 0$; the singularity itself is said to be strong
(weak) if all causal geodesics terminate in a strong (weak)
singularity.

According to a result of Tipler \cite{tipler}, a singularity will
be weak if $t^2R_{ab}(x(t))k^a(t)k^b(t)$ vanishes in the limit
$t\rightarrow 0$, where $k^a$ is tangent to the geodesic. Note
that here the Ricci tensor must be viewed as a function of the
parameter $t$, and not as a function of the space-time coordinates
$x^a$. Thus, if the Riemann tensor components, when twice
integrated {\em along the geodesic} give finite non-zero results,
the singularity must be weak. Some authors have concluded that the
same result holds if the Riemann tensor components, twice
integrated {\em as functions of the space-time coordinates}, yield
finite non-zero results. Thus a weak singularity has been deduced
from the existence of a regular $C^0$ metric
\cite{ori,ori-flan,HP}. We show here that this conclusion is not
valid by producing examples of space-times which contain {\em
strong}, regular $C^0$ singularities.

We consider a spherically symmetric space-time so that the line
element may be written using double-null coordinates as \be
ds^2&=&-2e^{-2f}dudv+r^2(u,v)d\Omega^2.\label{linel}\ee Take
$r=(v-u)/2$ to be the radius function of Minkowski space-time. The
gravitational strength of a non-central $(r>0)$ singularity is
completely determined by the limiting behaviour of $a(t)$, the
norm of an arbitrary radial Jacobi field orthogonal to a radial time-like
geodesic (here radial means lying in the 2-space spanned by
$\partial_u$ and $\partial_v$). We have shown elsewhere that the
tangential Jacobi fields orthogonal to arbitrary radial time-like
and null geodesics have finite and non-zero norm at a non-central
singularity. See \cite{ssss} for details. This function $a(t)$
obeys the {\em covariant} equation \be {\ddot
a}+2e^{2f}f_{uv}a&=&0, \label{main}\ee where the overdot indicates
differentiation along a time-like geodesic and the subscripts
indicate partial derivatives. The coefficient of $a$ is given in
terms of invariants by
\[ F(x^a):=e^{2f}f_{uv}=\frac{E}{r^3}+2\Psi_2-\frac{R}{12},\]
where $E$ is the Lema\^{\i}tre-Misner-Sharp mass, $\Psi_2$ is the
Newman-Penrose Weyl tensor term and $R$ is the Ricci scalar. If
$a(t)$ is finite and non-zero in the limit $t\rightarrow 0$ (the
location of the singularity), then this particular geodesic
terminates in a weak singularity. Otherwise the geodesic
terminates in a strong singularity. Thus from WKB analysis, if
\[\lim_{t\rightarrow 0}t^2F(t)\neq 0,\]
then the geodesic terminates in a strong singularity \cite{ssss},
where the $t$ dependence indicates that the space-time invariant
$F$ is evaluated along the geodesic.

The radial time-like geodesics of (\ref{linel}) satisfy \be {\dot
u}{\dot v}&=&\frac{1}{2}e^{2f},\label{geo1}\\ {\ddot u}-2f_u{\dot
u}^2&=&0,\label{geo2}\\ {\ddot v}-2f_v{\dot
v}^2&=&0.\label{geo3}\ee We consider a space-time with $f=f(x)$,
where $x:=u+v$. Then the geodesic equations reduce to the single
second order equation
\[ {\ddot x}-2f^\prime{\dot x}^2+2f^\prime e^{2f}=0\]
which has the first integral \be {\dot
x}^2&=&2e^{2f}(1+ce^{2f})=:y^2(t).\label{xdot}\ee To prove our
result, we show that there exists a solution of (\ref{xdot}) with
the following properties.
\newcounter{props}
\begin{list}
{(\roman{props})}{\usecounter{props}}
\item $x(0)=0$
\item $0<e^f(0)<+\infty$
\item $f_{uv}(x)\rightarrow\infty$ as $x\rightarrow 0$
\item $t^2f_{uv}(t)\not\rightarrow 0$ as $t\rightarrow 0^-$.
\end{list}

Property (iii) shows that there is a scalar curvature singularity
at $x=0$, which from the definition of $x$ and $r$ is non-central.
Properties (i) and (ii) indicate that the metric is regular at
the singularity and condition (iv) shows that the geodesic
terminates in a strong curvature singularity. We show how to
construct a $C^0$ metric from this solution below.

According to our ansatz, $f_{uv}=f^{\prime\prime}(x)$, and from
(\ref{xdot}), \be
e^{2f}(x(t))&=&-\frac{1}{2c}(1+(1+4cy^2(t))^{1/2}).\label{finy}\ee We
then calculate the derivative of $f$ using $f^\prime={\dot
x}^{-1}{\dot f}$; the second derivative obeys \be
f^{\prime\prime}e^{4f}&=&-y^{-1}z{\ddot
y}e^{2f}-4z^3{\dot y}^2-6z^2{\dot
y}^2,\label{fterm}\ee 
where $z=(1+4cy^2(t))^{-1/2}$. Now take \be y(t)=1+t\sin(\frac{1}{t})
\label{ydef}\ee for $t\neq 0$ and $y(0)=1$. Then $y$ is continuous
on ${\bf R}$, and so $x(t)$ is defined, is differentiable and
from (\ref{xdot}), is monotone in a neighbourhood of $t=0$. 
Thus the inverse function theorem
may be applied to find $t(x)$ in a neighbourhood of $x=0$, and
then (\ref{finy}) gives the metric function $f$ as a continuous
function of $x$, regular at $x=0$. The constant of integration may
be chosen so that $x(0)=0$. A straightforward calculation gives
the right hand side of (\ref{fterm}), and it is easily verified
that the limit $\lim_{t\rightarrow 0^-}t^2f^{\prime\prime}e^{2f}$
does not exist, and so as claimed, the singularity cannot be weak.
Note that the method of construction of the metric, while
somewhat complicated, guarantees that the singularity is
gravitationally strong along {\em every} radial time-like geodesic
running into it.

The key to constructing this example was to find 
$y(t)\in 
C^0(\alpha,0]\cap C^2(\alpha,0)$ such that the limit
$\lim_{t\rightarrow 0^-}t^2H(y,{\dot y},{\ddot y},t)$ is non-zero,
where $H$ is given by the right hand side of (\ref{fterm}), using
(\ref{finy}) for $f$. Given that continuous, non-differentiable
functions vastly outnumber differentiable ones, for the ansatz
used here, weak singularities would be the exception rather than
the rule even for a regular $C^0$ metric.

As mentioned above, this result shows that recent arguments that
the null CH singularity is gravitationally weak are incomplete.
This incomplete argument was used in analytical and numerical
studies of black holes by several authors,
\cite{ori,burko,burko-ori,HP} and also in a study of the generic
nature of null singularities \cite{ori-flan}. We do not claim that
these authors' results are wrong, but that the argument is flawed. In
fact in another study of the black hole interior, Brady {\em et
al} \cite{BDM} showed that in agreement with other authors the CH
singularity is $C^0$ and regular, and crucially, that the proper
time $t$ along timelike geodesics intersecting the singularity is
a smooth function of the coordinates used. These are sufficient
conditions to allow the use of Tipler's result \cite{tipler} and
to conclude that the singularity is weak. However in general, the
question of the gravitational strength of the CH singularity
remains open. It may be that the null nature, or some other feature, 
of the CH singularity enforces weakness; the singularity studied 
here is spacelike.

\end{document}